\documentclass{article}
\usepackage{graphicx}
\usepackage{url}

\begin{document}

\title{E-voting in Estonia}
\author{Dylan Clarke and Tarvi Martens \\Newcastle University, Estonian Electronic Voting Committee }
\date{}
\maketitle

Estonia\index{Estonia} has one of the most established e-voting systems in the world.
Internet voting - remote e-voting using the voter's own equipment
- was piloted in 2005 \cite{madise2006voting} (with the first real
elections using e-voting being conducted the same year) and has been
in use ever since. So far, the Estonian internet voting\index{Internet Voting} system has
been used for the whole country in three sets of local elections,
two European Parliament\index{European Parliament} elections and three parliamentary elections
\cite{NEC15Internet}. 

This chapter begins by exploring the voting system in Estonia\index{Estonia}; we
consider the organisation of the electoral system\index{Electoral System} in the three main
kinds of election (municipal, parliamentary and European Parliament\index{European Parliament}),
the traditional ways of voting and the methods used to tally votes
and elect candidates. Next we investigate the Estonian national ID
card, an identity document that plays a key part in enabling internet
voting to be possible in Estonia\index{Estonia}.

After considering these pre-requisites, we describe the current internet
voting system, including how it has evolved over time and the relatively
new verification mechanisms that are available to voters. Next we
discuss the assumptions and choices that have been made in the design
of this system and the challenges and criticism that it has received.
Finally, we conclude by discussing how the system has performed over
the 10 years it has been in use, and the impact it appears to have
had on voter turnout\index{Turnout} and satisfaction.

\section{Voting in Estonia\index{Estonia}}

The current Estonian system of governments started in 1991 when the
country declared independence from the Soviet Union. The Riigikogu
(Parliament of Estonia\index{Estonia}) consists of 101 members who are elected for
a period of four years \cite{OCSE11Estonia}. The electoral seats
are divided between 12 districts (shown for 2011 in Table \ref{tab:Electoral-Districts-Riigikogu}).
Candidates stand for a particular district and each voter casts one
vote for their favourite candidate among those standing in the district
in which they live. This vote can also have an indirect effect on
other districts due to the way in which electoral seats are allocated.
Seat allocation occurs in the following manner \cite{Rigiikogu02Election}:

Every party with candidates standing in more than one district must
submit an ordered listing of their candidates. This list indicates
their preference for the order in which seats awarded to the party
as a whole will be allocated (if there are seats left for which no
individual candidate has won outright). The allocation of seats takes
place in three rounds, with each round only proceeding if there remain
unallocated seats after the previous round. 

The first round takes place on a district level and is based on a
simple quota\index{Quota}. This simple quota\index{Quota} is calculated by dividing the number
of votes cast in the district by the number of seats in the district.
Any candidate who has received votes equal or greater than the simple
quota\index{Quota} is elected.

The second round also takes place on a district level and is based
on the number of votes allocated to each party in the district. The
candidates for each party are ordered by the number of votes they
received and the total number of votes for the party in this district
are calculated. If the total number of votes for the party is less
than 5\% of the votes cast then the party is awarded no further seats.
Otherwise, the total number of votes for the party is then divided
by the simple quota\index{Quota}, and the number of candidates from the party who
were elected in the first round is deducted. This gives the number
of candidates to be elected for each party in round two, and the actual
candidates to be elected are determined using the list of candidates
ordered by number of votes. 

The third round takes place on a national level and is again based
on the number of votes allocated to each party in Estonia\index{Estonia} as a whole.
A modified version of the d'Hondt method\index{D'Hondt Method} is used to allocate the remaining
seats between parties, and the ordered list that each party submitted
before the election is used to decide which candidates are elected
for each party.

The modified d'Hondt method\index{D'Hondt Method} works as follows: Each remaining seat
is considered in turn and a quotient is calculated for each party,
with the party with the highest quotient winning the seat. The quotient
is calculated by dividing the total number of votes for the party
by a function of the number of seats awarded to the party so far (in
the Estonian election the function of the number of seats is $f(n)=(1+n)^{0.9}$).
As with the second round, parties that have received less than 5\%
of the total votes cast are not allocated any seats, even if they
have the highest quotient at some point.

Estonia\index{Estonia} elects six candidates to the European Parliament\index{European Parliament} for a period
of five years. Each voter casts one vote for their favourite candidate
as with Riigikogu elections, but all candidates are elected nationally
rather than on a district basis. Each party submits an ordered list
of candidates, with independent candidates being treated as if they
were part of a candidate list with one candidate. European Parliament\index{European Parliament}
seats are allocated in the same way as for round three of Riigikogu
elections, except that the function of the number of seats already
allocated used in the d'Hondt method\index{D'Hondt Method} is $f(n)=1+n$.

Each city or rural municipality elects councillors every 4 years.
The number of councillors elected depends on the size of the municipality
and the methods used for allocating seats are as follows: For all
municipalities with one electoral district, seats are first allocated
in the same way as for round one in the Riigikogu elections, and remaining
seats are allocated in the same way as for the European Parliament\index{European Parliament}
elections. For municipalities with more than one electoral district,
a similar method is used to that in Riigikogu elections. Rounds one
and two are performed within districts and round three, if required,
is performed within the municipality as a whole. The d'Hondt method\index{D'Hondt Method}
used in round three uses the function $f(n)=1+n$ rather than the
function used in Riigikogu elections. 

We note that, while the allocation of seats in these elections is
relatively complex, the actual voting is very simple. Each voter has
only to chose their favourite candidate and cast one vote for that
candidate.

Voters are allowed to vote in a variety of ways, with the primary
methods being voting in person at polling stations, and internet voting\index{Internet Voting}
which we address in future sections. Voters who choose to vote in
person have the choice of either voting on election day or casting
an advance vote during an earlier period (initially three days but
extended to seven days from 2009 onwards) \cite{OCSE11Estonia}. This
advance vote can either be cast in their own voting district, or in
any other district at a designated place for outside-district voting.
Provision is also made in parliamentary and European Parliament\index{European Parliament} elections
for voters who are overseas to cast votes during the advance period. 

Advance voting takes place differently depending on whether the vote
is cast inside-district or outside-district. In both cases, the voter
enters a private booth and fills out a ballot paper. When voting inside-district,
the voter then folds the ballot paper, allows an offical to attach
a seal to it and deposits it in the ballot box. When voting outside-district
the voter places the ballot paper into an envelope, and then places
the envelope into a second envelope with their name, address and personal
identification number written on it. This outer envelope is then placed
in a ballot box for outside-district voters, to later be delivered
to the district in which the voter is registered.

There is some opportunity for those who have used internet or advance
voting to cancel out a vote with one that has higher precedence. Internet
votes can be cancelled by casting any type of physical advance vote,
and advance votes cast outside of the voter's district can be cancelled
by casting an advance vote inside the district.

It is not possible to cancel out an internet or advance vote by casting
a vote on election day, as polling stations will not allow those who
have used one of these mechanisms to cast a ballot.

Voters who are unable to make use of other voting methods due to ill
health or unexpected circumstances are allowed to apply to cast a
home vote on election day. This vote is cast in person when two election
officials visit their home to receive the ballot \cite{NEC15Voting,Rigiikogu02Election}.

\begin{table}
\protect\caption{Electoral Districts For Riigikogu Elections (2011) \label{tab:Electoral-Districts-Riigikogu}}

\centering{}%
\begin{tabular}{|c|c|c|}
\hline 
District Number & Area & Seats\tabularnewline
\hline 
\hline 
1 & Tallinn (Haabersti, P\~{o}hja-Tallinn and Kristiine districts)  & 9\tabularnewline
\hline 
2 & Tallinn (Kesklinn, Lasnam\"{a}e and Pirita districts) & 11\tabularnewline
\hline 
3 & Tallinn (Mustam\"{a}e and N\~{o}mme districts) & 8\tabularnewline
\hline 
4 & Harjumaa (exluding Tallinn) and Raplamaa & 14\tabularnewline
\hline 
5 & Hiiumaa, L\"{a}\"{a}nemaa and Saaremaa & 6\tabularnewline
\hline 
6 & L\"{a}\"{a}ne-Virumaa & 5\tabularnewline
\hline 
7 & Ida-Virumaa & 8\tabularnewline
\hline 
8 & J\"{a}rvamaa and Viljandimaa & 8\tabularnewline
\hline 
9 & J\~{o}gevamaa and Tartumaa (excluding Tartu) & 7\tabularnewline
\hline 
10 & Tartu & 8\tabularnewline
\hline 
11 & V\~{o}rumaa, Valgamaa and P\~{o}lvamaa & 9\tabularnewline
\hline 
12 & P\"{a}rnumaa & 8\tabularnewline
\hline 
\end{tabular}
\end{table}

\section{Estonian National ID Cards}

The Estonian national ID card was first introduced in 2002 \cite{martens2010electronic}.
It is a mandatory identity document for all Estonian citizens and
permanent residents over the age of 15. The document has the card
holder's full name, gender, national identification number, date of
birth, citizenship status, card number, card expiration date and photo
printed on it, and can be used as a primary travel document with the
European Union\index{European Union}. The card has also been used in the past as a ticket
for public transportation in major cities such as Tallinn \cite{eEstoniaID};
when a ticket was purchased the cardholder's national identification
number was stored in a central database and ticket inspectors could
then read the card with a handheld terminal and query the database
when they encountered the card holder.

More importantly from our point of view, the ID card contains a chip
with digital versions of the printed data, two 2048 bit RSA\index{RSA} key pairs
and certificates for these key pairs. The first key pair is used for
authentication and the certificate binds the citizen's public key
to their name, national identification number and a government issued
e-mail address. The second key pair is used for digital signing and
the certificate binds the citizen's public key to their name and national
identification number.

The chip is also capable of answering authentication challenges using
the first key pair, and generating digital signatures\index{Digital Signature} using the second
key pair, removing any need for the private keys to be communicated
outside of the chip. Interaction with the chip takes place using a
personal computer with a card reader and custom software.

The citizen is issued three PINs with the card which are used as follows:
PIN1 is required by the card before answering each authentication
challenge and PIN2 is required by the card before generating each
digital signature\index{Digital Signature}. If PIN1 or PIN2 is entered incorrectly three times
then the card is locked until the third longer PIN PUK is entered.

The national ID card is valid for 5 years, and after this point the
certificates expire. The certificates are also revoked if the owner
reports the card stolen or compromised.

Estonian citizens and permanent residents are also allowed to request
two other forms of digital identification: digi-ID and mobiil-ID.
Digi-ID is a card similar to the national ID card that is designed
only for online use. The digi-ID card does not have a printed photo
of the citizen, and contains less personal data then the national
ID card, while still providing the authentication and digital signature\index{Digital Signature}
functionality \cite{DigiID}. 

As of December 2014, Digi-ID has also become available to people who
are not citizens or permanent residents of Estonia\index{Estonia}. Possession of
a Digi-ID card allows a non-resident to make use of the Estonian authentication
and digital signature\index{Digital Signature} mechanisms online, without conferring any additional
rights with regard to Estonia\index{Estonia}. If the Digi-ID card becomes popular
outside Estonia\index{Estonia} then this has the potential to increase the investment
in the Estonian ID card system and the scrutiny to which it is subject,
a situation that can only be positive for an e-voting system relying
on the security of the ID card system.

Mobiil-ID provides similar functionality to digi-ID, but is built
into a mobile phone SIM card rather than a chip and PIN card. This
enables the citizen to perform digital authentication and signing
using their mobile phone with no extra hardware.

\section{The Internet Voting System }

The Estonian internet voting\index{Internet Voting} system is built around the simple and
well studied concept of public key cryptography\index{Public-Key Cryptography} \cite{NEC10General}.
Voters encrypt their ballots with the public key of the election system,
and then digitally sign them with their own private key. This private
key is the signing key on the voter's national ID card, which guarantees
that every voter will have a private key and a means of using it,
and that the election authority can reliably associate each voter
with their correct public key.

Simplicity is a concept that is heavily emphasised in the system design.
Rather than try to provide absolute end-to-end verifiability\index{Verifiability}\index{End-to-End Verifiability} and provable
anonymity and privacy, the designers instead attempted to produce
a system that can provide as much security as a paper based system,
while using simple components and well understood technologies. 

The system does not make use of a technology such as mix-nets\index{Mix-net / Mixnet} to protect
voter privacy. Instead, a simple separation of functions is used,
where no one piece of hardware or user ever holds both the server's
private key and a ballot with the digital signature\index{Digital Signature} attached. Vote
integrity is protected both by this separation of functions and by
the use of auditable logs produced by each component.

We begin by detailing the 7 main components of the system (shown in
Figure \ref{fig:System-Components}), we then provide further details
of the interplay between the components, and finally we describe how
the audit logs and the verification application can be used to check
certain properties of the system.

\begin{center}
\begin{figure}

\protect\caption{System Components \label{fig:System-Components}}

\includegraphics[scale=0.5]{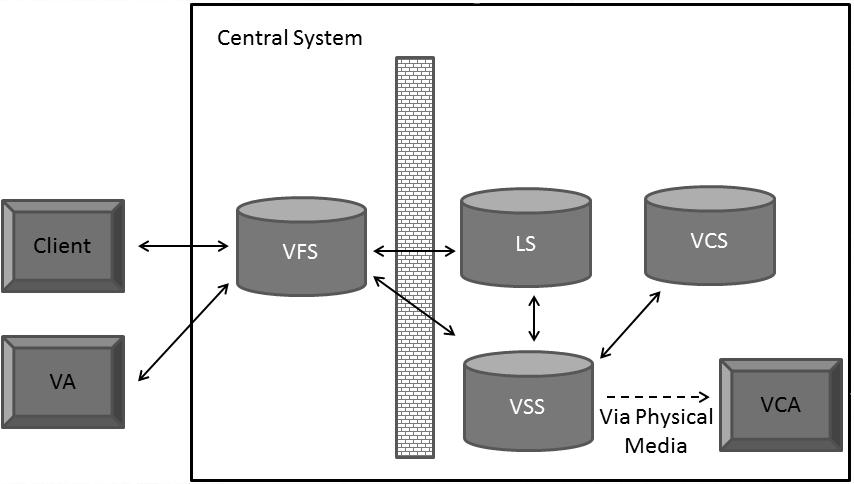}
\end{figure}

\par\end{center}

\subsection{System Components }
\begin{enumerate}
\item \textbf{The voting application (client):} The client is downloaded
to the voter's PC and allows the voter to communicate with the vote
forwarding server and cast a ballot. The client is available for Windows,
Mac OS and Linux.
\item \textbf{The verification application (VA):} The VA is an application
for smart phones and tablets that enables the voter to verify that
their ballot is held by the VSS. It is designed as a smart phone and
tablet application so that the voter will use two independent devices
during voting; a PC to cast the ballot and a smart phone or tablet
to verify that it is stored correctly. The verification protocol used
is explained in section \ref{sub:Auditing-and-Verification}.
\item \textbf{The vote forwarding server (VFS):} The VFS authenticates voters
and forwards requests between the client and other server-side components.
The VFS is the only server-side component that is publicly accessible.
It produces web server and application logs, but does not produce
logs of voting activity. The VFS also holds lists of all eligible
voters and candidates.
\item T\textbf{he log server (LS):} The log server is accessible by the
VFS and VSS. It stores log entries from these two servers. 
\item \textbf{The vote storage server (VSS):} The VSS receives votes from
the VFS and stores them if they are correctly signed. It is responsible
for communicating with the VCS to check signature validity, the VFS
to provide receipts to the client and providing a list of anonymised
ballots to the VCA. The VFS also handles the deletion of any ballots
belonging to voters who voted at a polling station and the logging
of all actions performed.
\item \textbf{The validity confirmation server (VCS):} The validity confirmation
server has the ability to check that signing certificates used to
create digital signatures\index{Digital Signature} are valid and to provide signed confirmation
attesting to this fact. The VCS is an external component, and is used
by all service providers making use of Estonian national ID cards
for authentication and/or digital signing.
\item \textbf{The vote counting application (VCA):} The vote counting application
is air-gapped from the rest of the internet voting\index{Internet Voting} system and is only
used after the voting period has concluded. It is responsible for
decrypting all of the valid votes and tallying them by constituency.
The VCA connects to an HSM that holds the private key corresponding
to the election public key. It also has the capability to store its
own logs, separate from those in the LS.
\end{enumerate}

\subsection{Normal System Operation}

Normal system operation involves the following steps:
\begin{enumerate}
\item The voter downloads the voting client.
\item The voter uses their national ID card (or digi-ID card) and card reader
to authenticate to the VFS. This requires the voter to enter PIN1
into the card reader in the same way that they would authenticate
for any other service with the ID card.
\item The client requests the list of candidates for the voter from the
VFS.
\item The VFS checks the voter list and determines the voter's voting region.
\item The VFS checks the candidate list and produces a list of all candidates
in the voter's voting region.
\item The VFS forwards the regional candidate list to the client.
\item The voter chooses a candidate to vote for.
\item The client generates an encrypted ballot for that candidate. The encrypted
ballot consists of the chosen candidate, encrypted
with the election public key. A random number generated by the client is also used in the encryption.
\item The client prompts the voter to sign the encrypted ballot.
\item The client uses their ID card to sign the encrypted ballot. This requires
the voter to enter PIN2 into the card reader.
\item The client sends the signed encrypted ballot to the VFS.
\item The VFS receives the signed encrypted ballot and forwards it to the
VSS. 
\item The VSS receives a signed encrypted ballot from the VFS, and contacts
the VCS to query whether the certificate used for signing is valid.
\item The VCS checks the certificate and informs the VSS. If the certificate
is valid then it provides signed validity confirmation. 
\item If the certificate or signature is invalid then the VSS rejects the
ballot and informs the VFS which in turn informs the client. The protocol
then stops at this point.
\item If the digital signature\index{Digital Signature} and the signing certificate are valid then
the VSS stores the encrypted ballot and also stores the voter's identification
number and a hash of the ballot in the log file LOG1
\item The VSS checks if the voter has already voted.
\item If the voter has already voted then the VSS stores a hash of the previous
ballot, the voter's identification number and the revocation reason
in LOG2 and deletes the previous ballot. 
\item The VSS sends a receipt to the VFS to be forwarded to the client.
At this point the voter can choose to follow the verification procedure
detailed in section \ref{sub:Auditing-and-Verification}.
\item When the voting period has ended, a list of e-voters is printed from
the VSS, for each polling station. Each list is sent to the corresponding
polling station and checked against the list of people who have voted.
If the voter has voted at the polling station then a request for cancelling
the e-vote is sent from the polling station. Any ballots from these
voters are deleted, with the voter's identification number, a hash
of the ballot and the reason for revocation being stored in LOG2.
\item The VSS sorts all ballots by constituency and removes their digital
signatures, storing these digital signatures\index{Digital Signature} as a proof of who has
voted. It then stores the hash of each vote in LOG3 (which is titled
``votes which go for counting''). 
\item The encrypted ballots without signatures are then exported onto physical
media and transferred to the VCA.
\item The VCA accepts, via physical media, the list of encrypted ballots
sorted by constituency..
\item A threshold number of election officials insert cryptosticks with
USB interfaces into the VCA.
\item This HSM decrypts all of the encrypted ballots.
\item The VCA then processes the decrypted ballots for each constituency
in turn. For each ballot, the VCA checks if the candidate chosen is
a valid candidate for the voter's constituency.
\item If the candidate is not valid then the vote is not counted and the
hash of the ballot is recorded in LOG4.
\item If the candidate is valid then the tally for that candidate is increased
by one and the hash of the ballot is stored in LOG5.
\end{enumerate}

\subsection{Auditing and Verification Capabilities \label{sub:Auditing-and-Verification}}

Immediately upon completion of tallying it is possible to audit the
consistency of the system by checking that the entries in LOG1 are
the same as the combined entries in LOG2 and LOG3 (i.e. that the ballots
accepted by the VFS are the same as the combination of the ballots
revoked by the VFS and the ballots sent for counting to the VCA) and
that the entries in LOG3 are equal to the combination of the entries
in LOG4 and LOG5 (i.e. that the ballots sent to the VCA are the same
as the combination of the invalid ballots and valid ballots). This
auditing is likely to detect system faults, but will not necessarily
detect malicious intruders as they may have the capability to determine
what is logged.

The verification application allows the voter to verify that the vote
they submitted is held on the vote storage server. This verification
proves the content of the vote to the voter, which may provide some
risk of coercion\index{Coercion}. However, this verification can only performed for
up to 30 minutes after the vote has been cast, can only be performed
three times, and does not leave the voter with a proof of vote. Remote
voting systems generally have a vulnerability\index{Vulnerability} to coercion\index{Coercion} as a coercer
could simply decide to present when the vote was cast; this verification
mechanism merely gives the coercer an extra 30 minute window in which
to be present. The voter also still has the option to re-vote. The
verification protocol is as follows :
\begin{enumerate}
\item When the voter casts a vote, the voting application generates a random
number that is used in the encryption of the ballot. 
\item When the encrypted vote is received, the vote forwarding server gives
the voting application a session code referring to this encrypted
vote.
\item The voting application generates a QR code containing the random number
and the session code.
\item The voter uses the verification application on a separate device to
scan this QR code.
\item The verification application contacts the vote forwarding server and
requests the encrypted vote corresponding to the session code.

\begin{enumerate}
\item If more than 30 minutes have elapsed from the vote being cast then
the vote forwarding server refuses to send the encrypted vote and
the protocol ends here.
\end{enumerate}
\item The verification application receives the encrypted vote and the list
of candidates in the voter's district.
\item The verification application use the random number to generate an encryption for each valid candidate until it finds an encryption
that matches the encrypted vote.
\item The verification application outputs the name of the matching candidate.
\end{enumerate}
We note that the requirement for the verification application to use
a brute force approach to discover the correct candidate prevents
the application from discovering which candidate the voter expects
to be contained in the encrypted vote. This prevents an attacker from
building a malicious verification application that simply returns
the candidate that the voter inputs. However, this does not prevent
more sophisticated attacks where a malicious verification application
communicates with a malicious voting application via some side channel\index{Side Channel}
to discover which candidate it should claim is encrypted. The verification
application also does not address the situation where a malicious
vote storage server provides a different encrypted vote to the verification
application from the one that is stored. 

However, as interesting as these theoretical attacks may be, in practice
they are unlikely to be possible without detection. If malicious verification
and voting applications are delivered widely to voters there is a
high chance of detection either due to a comparison between the malicious
application and the genuine application, or a malfunction caused due
to interoperability problems between malicious and genuine applications.
Similarly, attempts to modify the vote forwarding server or vote storage
server would have to circumvent many checks and security systems.

\section{Internet Voting Assumptions and Reception}

The Estonian National Electoral Committee produced a comprehensive
security and risk analysis for the Internet Voting system in 2003
and updated it in 2010 to include further concerns and developments
\cite{NEC10Analysis}. This analysis showed that the system had been
considered both as an electronic voting system and as a distributed
application, with the relevant security concerns for both highlighted.
Many of the issues discussed and conclusions reached were standard
in these fields, but we highlight some of the less usual assumptions.

``Universal verifiability\index{Verifiability}''\index{Universal Verifiability} was listed as a theoretical requirement as it was considered contradictory to the requirement of non-coercion\index{Coercion} in remote voting scenarios where it is assumed a coercer can see and influence any votes cast by the voter. Verifiability\index{Verifiability} in the style of ``recorded-as-intended'' was introduced later (in 2011) when the prevalence of smartphones allowed the introduction of the verification application.

The guarantee that votes will never become public (translated as absolute (fail-safe) of votes in the English version of the document) is also treated as theoretical given that unforeseen advances in cryptanalysis and computing power in general might lead to adversaries being able to break the cryptography used for the encryption of votes. Therefore further attention is paid to procedural controls like the handling and destruction of e-votes.

The Estonian Internet voting system has received considerable criticism
in the academic literature. This criticism covers both theoretical
and practical security concerns. The lack of traditional end-to-end verifiability\index{Verifiability}\index{End-to-End Verifiability}
causes voters to have to rely on secure hardware and carefully crafted
security procedures to ensure the integrity of the vote.

A relatively early criticism of the system was that using card readers
without their own display and keypad was a risk to the integrity of
the vote \cite{Schryen09Security}. This risk arises from the fact
that interaction with the ID card takes place through the same home
computer used for voting. If the computer is compromised then the
attacker can control both what is sent to the server and what is sent
to the ID card, with the voter having no other means to monitor the
process.

This threat may be partially mitigated by the more recent introduction
of the verification application, allowing the vote to be confirmed
on a second piece of hardware, and the increase in the number of voters
using card readers that feature an independent keypad.

The use of log files as a significant part of the system audit
has also been criticised \cite{Schryen09Security}. Log file disrepancies
are a good method for discovering system faults, but a malicious attacker
with control of one or more components may be able to control what
is logged in such a way that their modification of votes is undetectable.

More recently, the security of the system as used at the 2013 local
elections was criticised by Springall et al \cite{springall2014security}.
These criticisms were divided into four main areas: inadequate procedural
controls, lax operational security, insufficient transparency\index{Transparency} and
vulnerabilities\index{Vulnerability} in published code. The paper then went on to detail
client and server side attacks that the authors were able to mount
against their own reproduction of the Estonian internet voting\index{Internet Voting} system. 

The Estonian National Electoral Commission responded to the initial
reports of these criticisms with a statement in which they claimed
that all of the attack vectors discussed had already been accounted
for, it was not feasible to effectively conduct any of the attacks
to alter the results of the voting and that numerous safeguards were
in place to detect attacks and prevent tampering\index{Tampering} \cite{NEC14Comment}.
This statement also claimed that the website containing the criticisms
had numerous factual and detail errors, and did not sufficiently detail
the attacks. These alledged errors were not listed, so it is unclear
as to whether this claim also applies to the paper, or just the website
at that time.

Springall et al responded in turn and disputed all of the claims in
the Estonian National Electoral Commission's statement \cite{Springall14Response}.

The Internet Voting system has received considerably more approval
from the general public and the political parties of Estonia\index{Estonia}. The
2011 OCSE/ODIHR\index{Office for Democratic Institutions and Human Rights} Election Assessment Mission Report \cite{OCSE11Estonia}
claimed that ``Election stakeholders expressed confidence in the
overall process, including the Internet voting.''. It did however
also note that a legal challenge was made alledging a lack of reliability,
secrecy\index{Secrecy} and security of the Internet voting system. This complaint
was dismissed as unfounded and a challenge of this decision by the
Center Party was dismissed by the Supreme Court as it was not filed
in time. A further complaint was made alledging that some client names
were hidden under certain display settings in the voting application,
but this complaint was also dismissed as being without evidentiary
basis.

\section{System Performance}

The Estonian internet voting\index{Internet Voting} system has been used in eight major elections
over ten years; three Riikogu elections, two European Parliament\index{European Parliament} elections
and three local elections. The percentage of voters using the internet
voting system has grown over time for each of these three types of
elections \cite{NEC15Statistics}.

The first local election, which was also the first election in which
internet voting\index{Internet Voting} was used, had only 1.9\% of voters using internet
voting. This increased to 15.8\% for the second local election and
to 21.2\% for the third local election.

The first Riikogu election had 5.5\% of voters using internet voting\index{Internet Voting},
and this increased to 24.3\% for the second Riikogu election and 30.5\%
for the third Riikogu election. Simiarly, the first European Parliament\index{European Parliament}
election had 14.7\% of the voters using internet voting\index{Internet Voting}, while the
second had 31.3\% of voters using internet voting\index{Internet Voting}.

These figures show both that internet voting\index{Internet Voting} is used by a significant
percentage of voters in Estonia\index{Estonia} and that the system is capable of
handling large volumes of voters. The 24.3\% of voters in the 2011
Riikogu election equated to 140,846 people casting at least one internet
vote each.

This large scale usage of the internet voting\index{Internet Voting} systems has not highlighted
any observable major failures. The OCSE/ODIHR\index{Office for Democratic Institutions and Human Rights} EAM for the 2011 Riikokogu
elections did detail one minor fault; a single vote was found to have
been cast for an invalid candidate, something which should not have
been possible due to the design of the voting application \cite{OCSE11Estonia}. 

\bibliographystyle{plain}
\bibliography{bookref}

\end{document}